\pdfoutput=1
\documentclass[12pt, table]{article}

\usepackage{fancyhdr}
\usepackage[table,xcdraw]{xcolor}
\usepackage{natbib}

\usepackage[document]{ragged2e}
\usepackage{ulem}
\usepackage{graphics}
\usepackage[export]{adjustbox}
\usepackage{epsfig}
\usepackage{wrapfig}
\usepackage{shadow}
\usepackage{color}
\usepackage{dcolumn}	
\usepackage{amsmath}
\usepackage{amsfonts}
\usepackage{amssymb}
\usepackage{multirow}
\usepackage{enumitem}
\usepackage{graphicx}
\usepackage{booktabs}
\usepackage{lipsum}
\usepackage{siunitx}
\usepackage{gensymb}
\usepackage[final]{pdfpages}
\usepackage{pdflscape}
\usepackage{upgreek}
\usepackage[hidelinks]{hyperref}
\usepackage[many]{tcolorbox}
\usepackage[labelfont=bf,skip=0pt]{caption}
\usepackage{titlesec}
\usepackage[T1]{fontenc}
\usepackage{lmodern}

\usepackage{authblk}
\usepackage{orcidlink}

\titlespacing*{\section}
{0pt}{0.5\baselineskip}{0.5\baselineskip}
\titlespacing*{\subsection}
{0pt}{0.5\baselineskip}{0.5\baselineskip}
\titlespacing*{\subsubsection}
{0pt}{0.5\baselineskip}{0.5\baselineskip}

\definecolor{myred}{HTML}{FB6542}
\definecolor{dodgerblue}{RGB}{30, 144, 255}
\definecolor{myblue}{HTML}{375E97}
\definecolor{tableblue}{HTML}{1A73C9}
\definecolor{mygrey}{HTML}{363636}

\DeclareCaptionFont{myblue}{\color{myblue}}
\usepackage{sectsty}
\sectionfont{\color{myblue}}
\subsectionfont{\color{myblue}}

\newtcolorbox{bangbox}[2][]
{
  enhanced,
  colframe=white,
  title={#2},
  width=\textwidth, 
  colback=tableblue,
  coltext=white,
  center title,
  colbacktitle=myred,
  coltitle=white,
  fonttitle=\bfseries
}

\newtcolorbox{bangboxnh}[1][]
{
  enhanced,
  colframe=white,
  width=\textwidth, 
  colback=tableblue,
  coltext=white} 

\newcommand{\swri}
	{Southwest Research Institute, Boulder, CO}
\newcommand{\ncar}
	{National Center for Atmospheric Research, Boulder, CO}
\newcommand{\predsci}
    {Predictive Science Inc., San Diego, CA}
\newcommand{\umn}
    {University of Minnesota, Minneapolis, MN}
\newcommand{\amu}
    {American University, Washington, DC}
 \newcommand{\msfc}
    {NASA Marshall Space Flight Center, Huntsville, AL}
\newcommand{\cfa}
    {Center for Astrophysics $|$ Harvard \& Smithsonian, Cambridge, MA}
\newcommand{\calb}
    {University of California, Berkeley, CA}
\newcommand{\gsfc}
    {NASA Goddard Space Flight Center, Greenbelt, MD}

\usepackage[margin=1.0in, headheight=30pt]{geometry}
\begin{document}

\widowpenalty=0
\clubpenalty=0

\makeatletter
\renewcommand\Authfont{\fontsize{12}{14.4}\selectfont}
\renewcommand\Affilfont{\fontsize{9}{10.8}}
\renewcommand\AB@affilsepx{, \protect\Affilfont}
\makeatother


\title{\vspace{-3ex}\textbf{Magnetic Energy Powers the Corona: How We Can Understand its 3D Storage \& Release}}
\date{\vspace{-5ex}} 


\author[1]{Amir~Caspi%
\orcidlink{0000-0001-8702-8273}}
\author[1]{Daniel~B.~Seaton%
\orcidlink{0000-0002-0494-2025}}
\author[2]{Roberto~Casini%
\orcidlink{0000-0001-6990-513X}}
\author[3]{Cooper~Downs%
\orcidlink{0000-0003-1759-4354}}
\author[2]{Sarah~E.~Gibson%
\orcidlink{0000-0001-9831-2640}}
\author[2]{Holly~Gilbert%
\orcidlink{0000-0002-9985-7260}}
\author[4]{Lindsay~Glesener%
\orcidlink{0000-0001-7092-2703}}
\author[5]{Silvina~E.~Guidoni%
\orcidlink{0000-0003-1439-4218}}
\author[1]{J.~Marcus~Hughes%
\orcidlink{0000-0003-3410-7650}}
\author[6]{David~McKenzie%
\orcidlink{0000-0002-9921-7757}}
\author[1]{Joseph~Plowman%
\orcidlink{0000-0001-7016-7226}}
\author[7]{Katharine~K.~Reeves%
\orcidlink{0000-0002-6903-6832}}
\author[8]{Pascal~Saint-Hilaire%
\orcidlink{0000-0002-8283-4556}}
\author[9]{Albert~Y.~Shih%
\orcidlink{0000-0001-6874-2594}}
\author[1]{Matthew~J.~West%
\orcidlink{0000-0002-0631-2393}}

\affil[1]{\swri}
\affil[2]{\ncar}
\affil[3]{\predsci}
\affil[4]{\umn}
\affil[5]{\amu}
\affil[6]{\msfc}
\affil[7]{\cfa}
\affil[8]{\calb}
\affil[9]{\gsfc}
\maketitle
\vspace{-2ex}
\centering
\textbf{Additional Co-Authors:} Refer to attached spreadsheet.
\vspace{1ex}

\begin{figure}[h]
\centering
\includegraphics[width=0.9\columnwidth]{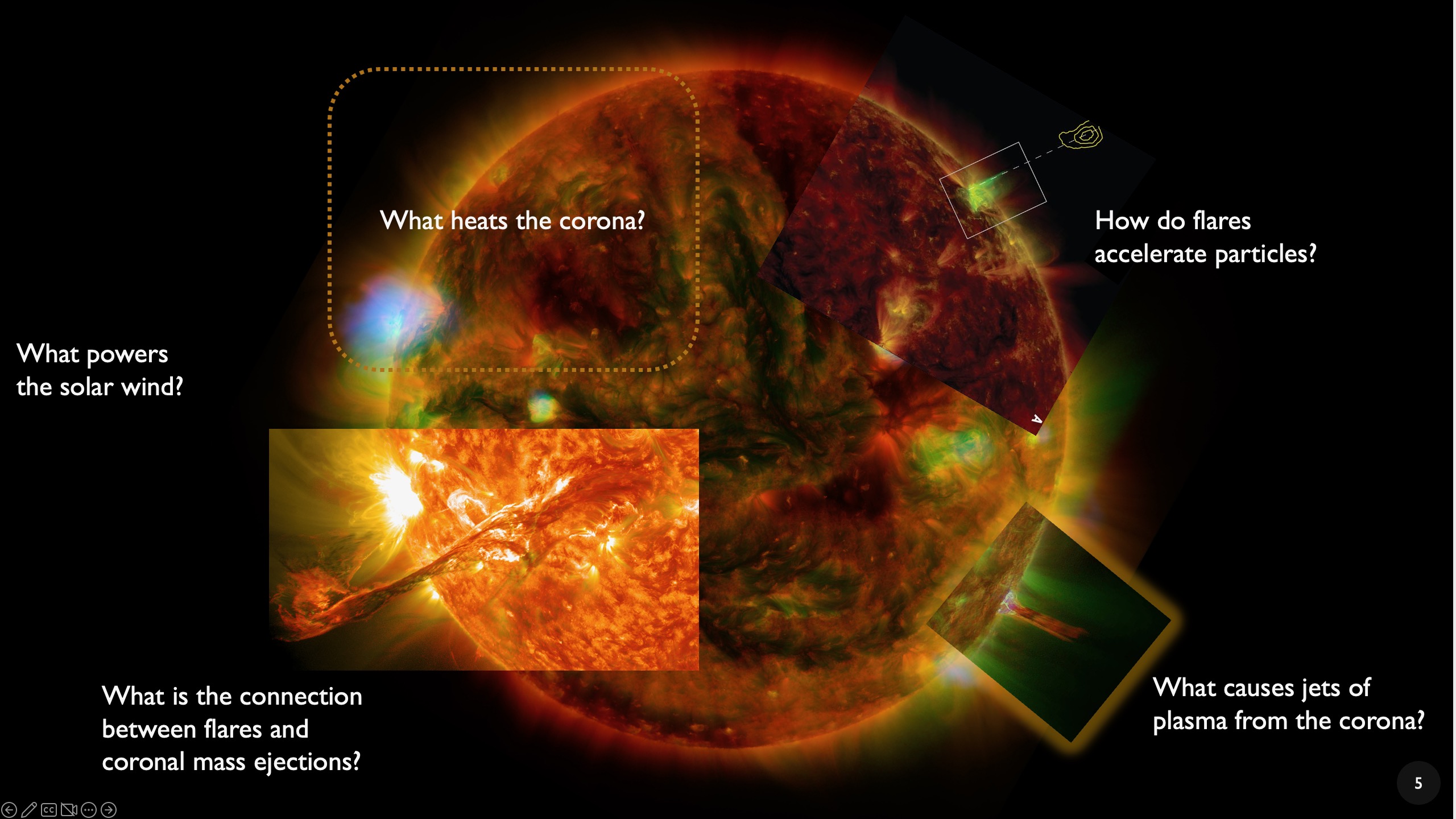}
\label{fig:overview}
\end{figure}

\begin{bangbox}{Synopsis}
\justifying
\noindent
The coronal magnetic field is the prime driver behind many as-yet unsolved mysteries: solar eruptions, coronal heating, and the solar wind, to name a few. It is, however, still poorly observed and understood. We highlight key questions related to magnetic energy storage, release, and transport in the solar corona, and their relationship to these important problems. We advocate for new and multi-point co-optimized measurements, sensitive to magnetic field and other plasma parameters, spanning from optical to $\gamma$-ray wavelengths, to bring closure to these long-standing and fundamental questions. We discuss how our approach can fully describe the 3D magnetic field, embedded plasma, particle energization, and their joint evolution to achieve these objectives.
\end{bangbox}

\fancypagestyle{firstpage}{%
  \chead{ 
  \begin{bangboxnh}
    \centering
    \textbf{A White Paper Submitted to the Solar \& Space Physics (Heliophysics) Decadal Survey --- 2024--2033} 
  \end{bangboxnh}
} }
\thispagestyle{firstpage}


\newpage
\cfoot{{\color{myblue} \thepage}}

\setcounter{page}{1}
\section{Introduction}
\label{sec:intro}\vspace{-2ex}
\justifying

Solar eruptions -- flares and their often-associated coronal mass ejections (CMEs) -- are the largest explosions in the solar system, releasing energies equivalent to 1--10 billion hydrogen bombs on timescales of seconds to minutes. This energy goes into heating solar coronal plasma to temperatures of 20--50 million Kelvin and into efficiently accelerating electrons to hundreds of MeV, protons/heavy ions to tens of GeV, and bulk flows of gigatons of atmospheric material into the heliosphere \citep{Flecher2011,Holman2011}. The photon and particle radiation from these eruptions comprises the most intense forms of space weather that can affect human assets in space (at Earth and other planets) and on the ground. Aside from eruptions, the Sun constantly releases significant energy, heating coronal plasma to 1--10\,MK, and drives a constant solar wind of energetic particles into the heliosphere \citep{Cranmer2019}. During the $\sim$150 years since these phenomena were first discovered \citep{Carrington1859}, it became clear that these processes are driven by the Sun's complex and dynamic magnetic field, which permeates the corona. Standard models of eruptions agree that coronal magnetic reconnection -- reconfiguration of the magnetic topology as discontinuous magnetic fields meet and annihilate -- is the process that frees magnetic energy stored in sheared/twisted magnetic fields \citep{Priest2002}. This reconnection and energy release are believed to occur within a current sheet separating sheared magnetic fields, but in the absence of direct coronal magnetic field measurements, the existence of this sheet and the nature of precursor energized fields can only be inferred (Fig.~\ref{fig:overview}).

Coronal magnetic fields are generally estimated from extrapolations that use surface magnetic field observations as their lower boundary condition, under restrictive assumptions (potential, nonlinear force-free, etc.) Estimates of critical properties of pre-eruption coronal magnetic fields, such as pre-existing flux ropes or the magnetic shear angle of coronal current sheets, are thus approximations. Moreover, standard eruptive models are generally phenomenological, describing the overall magnetic and plasma evolution rather than the detailed physics at critical interfaces, particularly at the reconnection site(s). Fig.~\ref{fig:overview} shows a cartoon of one of the most common standard flare/CME models \citep{Shibata1996} compared with extreme ultraviolet (EUV) and X-ray observations \citep[Zhu. 2017;][]{Seaton2018} and a numerical magnetic-field model \citep{Chen2020} for one of the best examples of an eruptive solar flare in recent years, highlighting both the general agreement and specific discrepancies between the model and observations.

Overall, there is little consensus in the literature on the underlying physical mechanisms that superheat plasma and accelerate particles and bulk flows, nor on the runaway instabilities that initiate reconnection to drive these processes. It is still poorly understood whether the hottest (20--50~MK) flare plasmas are heated directly from reconnection-associated processes, from shocks, or from collisions by accelerated particles \citep[e.g.,][]{Caspi2010,Longcope2011,Caspi2014}. Further, it is still hotly debated whether electrons and ions in solar flares are accelerated through stochastic processes, betatron acceleration, or wave-particle interactions, and exactly where in the flare this occurs \citep[e.g.,][]{Zharkova2011,Knuth2020}. Similarly, it is not known whether CMEs  are initiated by ideal or resistive magnetohydrodynamic (MHD) instabilities, and their pre-eruptive configuration has been extensively debated \citep{Klimchuk2001,Forbes2006,Green2018,Gibson2022}.

The discovery that the corona is millions of degrees was made more than 80 years ago \citep{Grotrian1939,Edlen1943}, and the link between this and solar wind outflow realized some 60 years ago \citep{Parker1959}. 
Still, no clear consensus has emerged for how the corona is heated, nor for how the solar wind is subsequently accelerated -- whether they are driven by impulsive flare-like processes, wave dissipation, or other mechanisms is still debated \citep{Klimchuk2006}.

The energetic processes behind flares and eruptions are hypothesized to extend down to small, unobservable scales, quasi-steadily releasing stored magnetic energy to heat coronal plasma. These same bursts could also accelerate small jets of particles that contribute to solar wind outflows \citep[e.g.,][]{Yang2013}. Such impulsive energy releases may themselves generate waves that further contribute to heating and solar wind acceleration \citep{vanBallegooijen2011}.

The coronal magnetic field is the only source of energy sufficient to power these processes across the gamut of spatial and energetic scales. Furthermore, the magnetic field and plasma are intimately linked and feed back upon one another during these phenomena. Despite major investments over the past 30--40 years, \textbf{{\color{myblue}it remains unclear exactly how the Sun’s magnetic field mediates nearly all energy storage, release, and transport to initiate and drive these plasma processes,}} and these long-standing fundamental questions remain open.

\begin{figure}
\centering
\includegraphics[width=0.98\columnwidth]{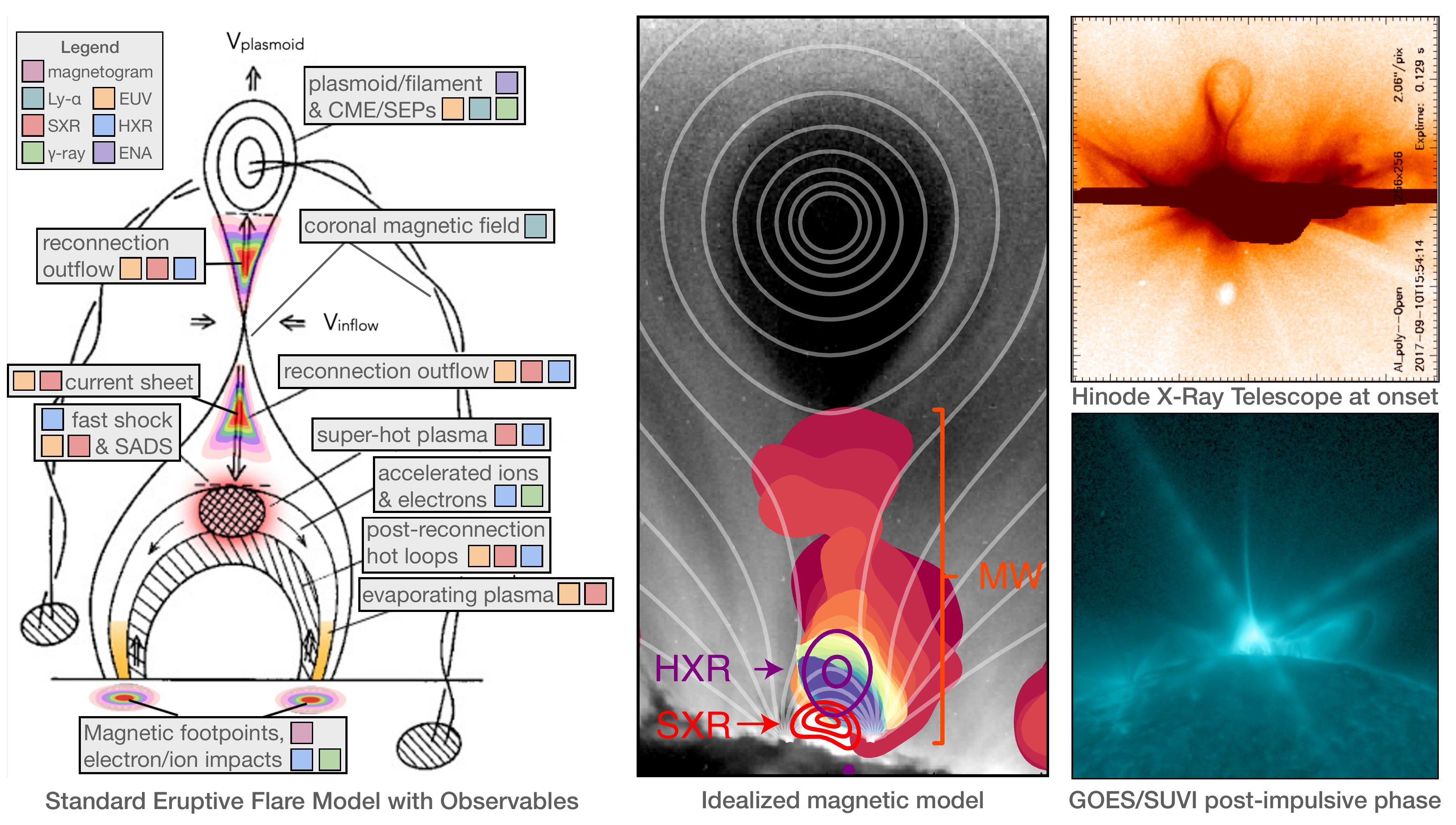}
\caption{\small \justifying Phenomenology and observables for eruptive solar flares \citep[left; adapted from][]{Shibata1996} compared to actual observations and numerical magnetic field model (middle: \citealt{Chen2020}; top right: Zhu 2017; bottom right: \citealt{Seaton2018}) highlighting the need for multiple integrated observables and multiple perspectives to probe the complete physics of magnetically-driven eruptions.}

\vspace{-4ex}
\label{fig:overview}
\end{figure}

Closing these questions has been hindered by two critical limitations. First, we lack a three-dimensional understanding of the energy release processes within the context of a 3D magnetic field. Remote-sensing observations from a single viewpoint are inherently 2D, providing information only in the plane of sky and integrating all emission along the line of sight. Recent simulations suggest that our assumptions about how 3D coronal structures can be extracted from 2D images with a single line of sight may be wrong \citep[the ``coronal veil'';][]{Malanushenko2022}. Additionally, the energy release manifests in highly localized regions and depends strongly on the 3D magnetic topology, so fully constraining models of these magnetically-driven processes inherently requires a 3D understanding. If a single 2D projection intrinsically fails to capture 3D information, then accurate 3D models are essentially impossible without additional measurements.

To highlight the importance of 3D understanding, consider: the MHD instabilities that trigger many eruptions depend strongly on the 3D topology of the erupting magnetic structure \citep{Zuccarello2014}. Similarly, localized brightenings can only be distinguished from line-of-sight projection effects using multiple viewpoints \citep{Guidoni2015}. Limited 3D reconstructions can be achieved by combining multiple viewpoints, as demonstrated by STEREO \citep{Howard2008} using epipolar projection of selected points to reconcile one viewpoint to another \citep[e.g.,][]{Aschwanden2008,Mierla2013,Dhuys2017}. However, these isolated projections cannot reconstruct the full 3D volume. While Solar Orbiter and Parker Solar Probe will provide additional wavelengths (visible light, EUV, and X-rays) to provide deeper insight into plasma energetics, these will have similar limitations \citep[e.g.,][]{Rochus2020}. 

More importantly, the ``coronal veil'' problem, if correct, undermines the viability of epipolar reconstruction purely from optically-thin emission. More advanced 3D reconstruction techniques such as tomography have been developed to reconstruct entire volumes, and do not suffer from the limitations of epipolar projection. But, with a single viewpoint, these rely on slow solar rotation \citep{frazin2005,kramar2016} or spacecraft orbital motion \citep[e.g.,][]{Vasquez2019} to obtain sufficient view angles and are not suitable for dynamic, or even slowly-evolving, processes. Combining the optically-thin emission from multiple lines of sight with additional constraints, such as the embedded coronal magnetic field, significantly improves the quality of 3D reconstructions. Recent techniques \citep[e.g., CROBAR;][]{Plowman2021} use magnetic models to constrain the locations of energized plasma, providing high-quality 3D reconstructions with many fewer lines of sight. These techniques currently use extrapolations from photospheric magnetic fields, which are not systematically available off the Sun-Earth line, limiting the regions where these reconstructions are viable. Moreover, magnetic field extrapolations are underconstrained and rely on \textit{a priori} assumptions, which cannot be validated without direct coronal diagnostics.

This highlights the second fundamental limitation, that we presently cannot measure the coronal magnetic field in regions relevant to these processes. Photospheric magnetic field measurements were first made 100 years ago and are now routine from space (SOHO/MDI, \citealt{Scherrer1995}; and SDO/HMI, \citealt{Scherrer2012}) and ground \citep[GONG;][]{Harvey1996Sci}. These measurements exploit the Zeeman effect, which imposes characteristic polarization signatures on spectral lines in the presence of an ambient magnetic field. However, such measurements are more difficult in the corona because of low densities, weak fields, and high temperature. While coronal fields can be extrapolated from photospheric boundary conditions \citep[e.g.,][]{Mikic2018}, direct measurements sensitive to coronal magnetic fields are required to validate reconstructions. 

Scattering polarization measurements of forbidden coronal lines using the Hanle effect in the infrared (IR) provide a direct measurement of the coronal magnetic field direction on the plane-of-sky \citep[CoMP;][]{Tomczyk2008}. In IR, the Hanle effect saturates at a few $\mu$G, and the only field strength diagnostic is the weak circular polarization of the Zeeman effect. While the vector field of well localized, largely static, weak-field, global magnetic structures at large heights above the solar surface can in principle be diagnosed \citep{Plowman2014,Dima2020,Paraschiv2022}, large telescopes (1-m class) are needed to measure the Zeeman circular polarization with sufficient precision \citep{Tomczyk2022}. Currently, only the plane-of-sky component of the coronal magnetic field can be inferred and only via indirect methods \citep{Yang2020}. 

A complementary diagnostic relies on the modification of scattering polarization of permitted lines by the coronal field through the unsaturated Hanle effect. In the corona, a particularly notable such line is the hydrogen Lyman-$\alpha$ line at 1216\,\AA, the brightest coronal emission in the solar spectrum, which can be used to determine magnetic field strengths in lower closed-field regions ($\sim$5--200 G) from which solar eruptions originate \citep{Raouafi2016,zhao2019,Supriya2021}. 
Such measurements are crucial to making breakthrough progress to close this knowledge gap and accessible even to a 0.1-m class instrument  \citep{casini2022}.

\begin{bangboxnh}
\justifying
\noindent Systematic multiwavelength observations including coronal magnetometry from multiple viewpoints are required to measure the location and extent of energy release within a dynamic, 3D coronal magnetic field \citep{Caspi2019,Seaton2021}.
\end{bangboxnh}

\vspace{-4ex}

\section{The Big Open Questions}
\label{sec:big-qs}\vspace{-2ex}

Determining the physical mechanisms that initiate magnetic reconnection to release stored coronal magnetic energy requires a broad 3D understanding of the field and plasma properties. New multi-perspective, multi-messenger observations are imperative. We discuss two categories of critical science questions addressed by such new observations.

\vspace{-2ex}

\subsection{ Magnetic Energy Storage \& Release}

\subsubsection{In 3D, where, how, and how much magnetic energy is stored prior to an impulsive event; and what magnetic configurations determine the timing, location, and extent of free energy release? What are the signatures of this release?}

Solar eruptions are powered by the energy stored in the coronal magnetic field via non-potentiality introduced by photospheric surface convection or emergence of twisted flux ropes \citep{Forbes1982}.
With sufficient stress, resistive \citep[e.g.,][]{Antiochos1999,Lin2000,Moore2001} or ideal \citep[e.g.,][]{Torok2005,FanGibson2007} instabilities trigger runaway rapid conversion of stored energy to other forms. Estimates of the stored energy depend entirely on knowledge of the coronal field, which is currently limited to inaccurate extrapolations. Which of these instabilities dominates energy release in specific circumstances is also hotly debated. Moreover, how these instabilities propagate and how much free energy they can efficiently convert is also poorly constrained \citep{Emslie2012,Aschwanden2017}.

Comprehensive 3D observations of coronal magnetic and plasma properties enable direct differentiation between eruption initiation models. Each instability is characterized by specific locations, timing, and mechanisms of energy release events within the 3D magnetic field \citep{Gibson2022}. A careful accounting of energization events including plasma heating, particle acceleration, and bulk flows provides the critical plasma properties needed to differentiate between these models. Such an accounting also provides an accurate determination of the amount and partition of free magnetic energy converted into various forms.

\vspace{-2ex}

\subsubsection{In 3D, where, how, and how much is magnetic energy released to drive coronal heating and solar wind outflow and what scaling laws relate small-scale impulsive or dissipative release events to major flare/eruptive ones?}

Since the discovery that the ambient corona is hot everywhere, a fundamental question is what process is responsible for extracting stored magnetic energy to ubiquitously heat the corona and, in turn, provide the energy to accelerate solar wind outflow \citep[see][]{Klimchuk2006}. \citet{Parker1988} argued that the same processes that give rise to heating in large flares could be responsible for heating the corona on very small scales through undetectable impulsive \textit{nanoflares}. Competing theories \citep{Kuperus1969} hold that heating instead results from dissipation of wave energy, such as from acoustic waves generated by photospheric convective motion. 

This question remains totally unresolved, but clues are abundant. For example, while magnetic-field footpoint motion causes the constant build-up of twist and shear, the field remains overwhelmingly smooth \citep{Deforest2007}, albeit complex in structure. The processes that ``comb'' the coronal field, and the extent to which they achieve this, are likely to be the same ones that extract energy from the field. Unfortunately both direct \citep[e.g.,][]{Rachmeler2014} and indirect \citep[e.g.,][]{Aschwanden2015} coronal field diagnostics are inadequate to characterize its complex structure \citep{Malanushenko2022} and to constrain models needed to understand how energy is both transferred into and extracted from the field over long timescales. New observations and reconstructions of the magnetic field and embedded energy release signatures \citep{SeatonWP2022} will overcome limitations baked into traditional methods to allow more accurate assessment of the structure of the coronal field and differentiation between competing models of coronal heating -- even from unresolvable events -- through the specific properties (e.g., temperature distribution) they imbue into the coronal plasma.

\vspace{-2ex}
\subsection{Energy Release Effects \& Transport}

\subsubsection{In 3D, where and how is plasma heated and particles accelerated before, during, and after eruptions? How do these processes vary between ions and electrons?}

Numerous observational signatures -- from hot candle-flame shaped structures \citep{Guidoni2015}, shrinking magnetic loops \citep[e.g.,][]{Reeves2008} and inflows \citep[e.g.,][]{Savage2012} when viewed on the limb; to the expansion of bright chromospheric flare ribbons when viewed from above -- support the standard flare reconnection model (see Fig.~\ref{fig:overview}). However, the coronal magnetic field, necessary to quantitatively validate models, must be inferred from indirect diagnostics \citep{Rachmeler2014,Chen2020b,Chen2020}. Hence, establishing a direct link between magnetic reconnection and subsequent manifestations of released energy such as super-hot coronal plasma and high-energy accelerated particles, remains infeasible for most events.

Significant circumstantial evidence suggests that intimate links exist. Heating of chromospheric plasma through energy deposition by accelerated particles has long been assumed \citep{Holman2011}. Heating directly in the corona is also strongly suggested -- most evident in super-hot ($\gtrsim$30~MK) looptop plasmas, with impulsive time profiles and higher altitudes than cooler plasma \citep{Caspi2010,Caspi2014,Caspi2015,Guidoni2015,Warmuth2016}. Many possible mechanisms have been posited, including ohmic heating \citep[e.g.,][]{Somov2000, Seaton2009, Caspi2010}; energy deposition by accelerated electrons \citep[e.g.,][]{Glesener2013,Caspi2015}; adiabatic or other compression \citep[e.g.,][]{Longcope2011,Reeves2017,Reeves2019}; and suppression of conductive cooling \citep{Bian2016}. Similarly, many electron acceleration models exist, including first-order \citep[magnetic island:][]{30} and second-order \citep[stochastic:][]{31,32,33,34} Fermi acceleration, electric fields \citep{35,36}, collapsing magnetic traps \citep{38,39,37}, and shocks \citep{40,41,42}. Many of these mechanisms are closely related to those for heating.

In both cases, no definitive consensus has emerged due to the lack of magnetic context and difficulties in comprehensive measurements of the 3D heated and accelerated electron populations, and their evolution in space and time. Ions are even more poorly understood, with both similarities and differences to electron behavior \citep[e.g.,][]{Hurford2003,Shih2009}, but much weaker observational signatures. Thorough understanding of these particle populations requires comprehensive and coordinated measurements of their observational signatures (EUV, X-rays, $\gamma$-rays, and energetic neutral atoms, all observed from space, ideally incorporated with ground-based microwave observations), including multiple perspectives to break line-of-sight degeneracies, to quantify directional differences in emission tied to acceleration mechanisms, and to place the locations of energization in context within the 3D magnetic field that powers these processes.

\vspace{-2ex}

\subsubsection{What are the properties of unresolvable energy release events that drive coronal heating and solar wind outflows? How do these relate to larger events?}

Energy release processes observable by current instruments have, so far, been shown to be inadequate to explain the heating required to maintain the hot corona, both globally and more locally within active regions. Parker's nanoflare hypothesis suggests that energy release events are governed by straightforward scaling laws. If the event frequency distribution is a power-law with slope $>$2, unresolvable events could deposit enough energy to account for coronal heating \citep{Hudson1991}. Many studies have attempted to characterize how energy release processes scale from larger flares to small ones \citep[e.g.,][]{Christe2008,Hannah2008,Kobelski2014,Glesener2020,Duncan2021,Cooper2021}, and have found significant correlations but with a widely disparate range of parameters \citep{Aschwanden2022}. Even the distribution function itself remains in question \citep{Verbeeck2019}, and there are hints of lower particle acceleration efficiency associated with smaller energy releases \citep[e.g.,][]{Battaglia2005,Hannah2008,Warmuth2016}.

Statistical studies have not closed these questions. Characterizing energy release processes down to unresolvable scales requires a systematic accounting of the various outlets for such release, including heating (temperature, density), particle acceleration, and bulk flows including jets and solar wind outflows. Such accounting is only possible with broad spectroscopic sampling of emission signatures of these processes, from EUV to X-rays \citep{Oka2022,Glesener2022} and $\gamma$-rays \citep{Shih2022}, and ideally incorporating ground-based microwave data \citep{Gary2022c,Chen2022a,Gary2022a}, to adequately measure the inherently broad temperature distribution and acceleration diagnostics. As with larger events, the 3D geometry -- both for the energy release and underlying magnetic field -- provides the context necessary to understand available free energy and how and where it can be released and transported even at small scales \citep{melrose1997,bale2022}. Multiple perspectives are therefore required, and, because these events extend to unresolvable scales, observations must be paired with data-constrained, physics-based models that can test predictions against reconstructed 3D plasma properties.

\vspace{1ex}

\begin{figure}[hb!]
\centering
\includegraphics[width=0.98\columnwidth]{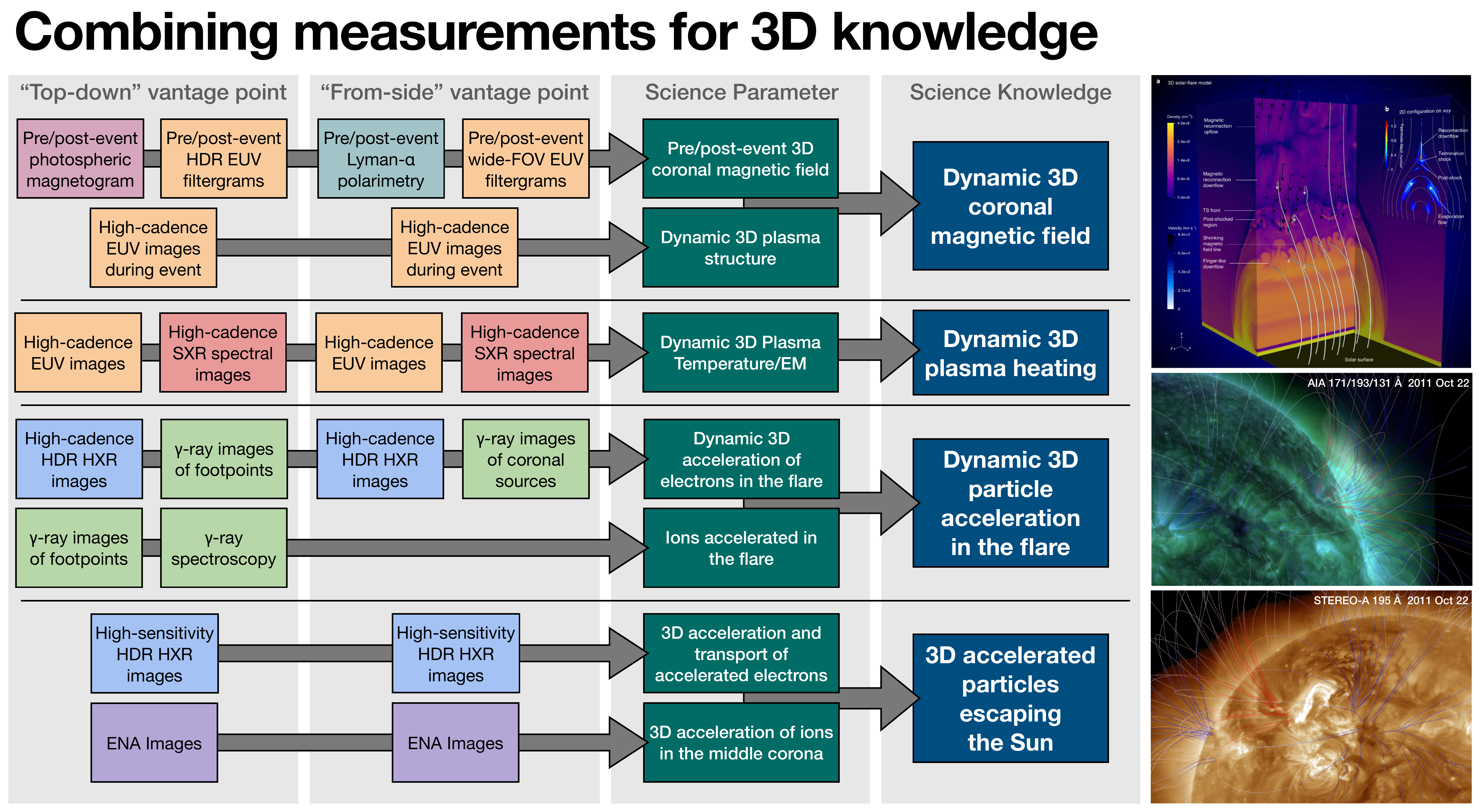}
\caption{\small \justifying Purposefully co-optimized multi-wavelength, multi-perspective observations combine with physics-based, data-constrained forward models to provide insight into underlying 3D physical parameters and driving  mechanisms to comprehensively close long-standing science questions (Sec.~\ref{sec:big-qs}, same color code as in the legends of the left panel of Fig.\ \ref{fig:overview}). The images (right) highlight how COMPLETE-like multi-wavelength measurements can be combined, through the flow at left, to yield a comprehensive physical understanding of events \citep[top right;][]{Shen2022}.
} 
\label{fig:flow}
\end{figure}

\section{The Need for Action}
\label{sec:action}\vspace{-2ex}

Many decades of innovation and investment in technical capability have yielded significant progress towards closing the above big questions and have proven that all of the necessary observations to achieve closure are indeed feasible. Every type of measurement required, from $\gamma$-rays to Lyman-$\alpha$, spectroscopy, polarization, imaging, and magnetography has been demonstrated in some fashion. Photospheric Doppler magnetography and narrowband EUV imaging are routine and performed operationally. Soft X-ray measurements are or have been routine in some respects (e.g., photometry via GOES XRS; spectroscopy via Hitomi, MinXSS, etc.; filter imaging via Yohkoh/SXT, GOES/SXI, Hinode/XRT, etc.). Hard X-ray and $\gamma$-ray imaging and spectroscopy were also done routinely by RHESSI, albeit with limited dynamic range due to a lack of focusing optics, and occasionally by NuSTAR, although only for small events as it was not designed for solar flare fluxes. Coronal magnetography is more recent but has been routinely made for weak fields by CoMP/UCoMP, and soon by DKIST. Energetic neutral atoms have been measured from Earth and planetary magnetospheres, though not yet applied to solar observations.

The benefits of solar observations from multiple perspectives have been demonstrated in some of the above regimes, first through STEREO and, more recently, by Solar Orbiter. Analyses using these missions, along with Earth-perspective observations such as from SDO, have laid the foundation for 3D reconstructions using optically thin emission (e.g., EUV). However, these missions were not designed to make the full suite of comprehensive measurements needed to close the questions described above. For example, STEREO did not include a magnetograph, while Solar Orbiter lacks soft X-ray diagnostics and is limited in its hard X-ray imaging capabilities. Neither mission can provide direct coronal magnetic field diagnostics. While missions to L5 have been conceived and are (in principle) planned, these focus primarily on space weather applications rather than the fundamental physics studies needed to close the above questions.

Recently, various aspects of the necessary measurements have been or will be proposed through individual missions, including multi-wavelength imaging spectroscopy \citep{Christe2022,Shih2022F}; global photospheric magnetography \citep{RaouafiWP2022}; and single-perspective coronal magnetography in strong-field regions relevant to these processes (Casini et al. 2022), as well as flare and active region magnetography in microwaves \citep{Gary2022c,Gary2022b}. Each of these missions would yield significant new progress into various aspects of these science questions. Full closure, however, requires a level of coordination and co-optimization between measurements that cannot be achieved through independent missions, even if they were all to fly or operate at the same time, or through the current PI-led mission paradigm.

A new flagship mission concept -- COMPLETE \citep{CaspiWP2002_Mission} -- provides the unified approach necessary to achieve these objectives and provide closure on the above science questions. COMPLETE implements two instrument suites -- a broadband spectroscopic imager spanning from EUV, through X-rays, to $\gamma$-rays and energetic neutral atoms; and a comprehensive 3D vector surface and coronal magnetograph -- distributed across multiple spacecraft at the Earth-Sun Lagrange points L1, L4, and L5. The specific instrument complement on each spacecraft is intentionally optimized for the observations from its unique vantage point, and the instruments are designed from inception to work together as a single observatory. Data from all instruments is assimilated into a unified data analysis framework (Fig.~\ref{fig:flow}) to drive data-constrained physical models that reconstruct the 3D coronal magnetic and plasma properties in ways not achievable without deep integration strategies \citep{SeatonWP2022}. With a 5-year prime mission around the maximum of Solar Cycle 26, COMPLETE's co-optimized instrument suite for broadband imaging spectroscopy and 3D magnetography, multiple perspectives, and integrated data and modeling framework will finally provide the necessary, comprehensive measurements and analysis to close these science questions that have remained open for 60 to 160 years.

\newpage

\setcounter{page}{1}
\cfoot{{\color{myblue}References -- \thepage}}

\bibliography{references.bib}
\bibliographystyle{aasjournal}

\end{document}